\documentclass[fleqn,usenatbib]{mnras}

\usepackage{newtxtext,newtxmath}

\usepackage{atbegshi}
\usepackage[T1]{fontenc}
\usepackage{ae,aecompl}

\usepackage{graphicx}    
\usepackage{amsmath}    
\usepackage{amssymb}    

\usepackage{booktabs}

\newcommand{\mass}{$M_\mathrm{200m}$ }
\newcommand{\mar}{$\Gamma_\mathrm{dyn}$ }

\newcommand{\rsp}{$R_\mathrm{sp}$ }

\title[Splashback radius detection]{How Accurately Can We Detect the Splashback Radius of Dark Matter Halos and its Correlation With Accretion Rate?}

\author[Xhakaj et al.]{Enia Xhakaj$^{1}$\thanks{E-mail: exhakaj@ucsc.edu},
    Benedikt Diemer$^{2}$, Alexie Leauthaud$^{1}$, Asher Wasserman$^{1}$, 
    \newauthor
    Song Huang$^{1}$, Yifei Luo$^{1}$, Susmita Adhikari$^{3}$, Sukhdeep Singh$^{4}$
\\
\\
$^{1}$Department of Astronomy and Astrophysics, University of California, Santa Cruz, 1156 High Street, Santa Cruz, CA 95064 USA \\
$^{2}$NHFP Einstein Fellow, Department of Astronomy, University of Maryland, College Park, MD 20742, USA;\\
$^{3}$Kavli Institute for Particle Astrophysics and Cosmology, Stanford University, 452 Lomita Mall Stanford, CA 94305, USA \\
$^{4}$Berkeley Center for Cosmological Physics, University of California, Berkeley, Berkeley, CA, USA 
}

\date{Accepted XXX. Received YYY; in original form ZZZ}

\pubyear{2019}

\begin{document}
\label{firstpage}
\pagerange{\pageref{firstpage}--\pageref{lastpage}}
\maketitle

\begin{abstract}
The splashback radius ($R_{\rm sp}$) of dark matter halos has recently been detected using weak gravitational lensing and cross-correlations with galaxies. However, different methods have been used to measure $R_{\rm sp}$ and to assess the significance of its detection. In this paper, we use simulations to study the precision and accuracy to which we can detect the splashback radius with 3D density,  3D subhalo, and weak lensing profiles. We study how well various methods and tracers recover $R_{\rm sp}$ by comparing it with the value measured directly from particle dynamics. We show that estimates of $R_{\rm sp}$ from density and subhalo profiles correspond to different percentiles of the underlying $R_{\rm sp}$ distribution of particle orbits. At low accretion rates, a second caustic appears and can bias results. Finally, we show that upcoming lensing surveys may be able to constrain the splashback-accretion rate relation directly.
\end{abstract}


\section{Introduction}

In the standard $\Lambda\mathrm{CDM}$ universe, dark matter halos form hierarchically due to the collapse of dark mark matter overdensities. This process can be described by the self-similar spherical collapse model, in which overdensities are considered to be composed of infinitesimally thin mass shells. These shells expand due to the Hubble flow, decelerate, start collapsing gravitationally and eventually virialize \citep{Fillmore1984Self-similarUniverse, Bertschinger1985Self-similarUniverse}. The boundary between the virialized and infalling shells is known as the splashback radius. It is defined as the radius where dark matter particles reach the apocenter of their first orbit as they accrete onto dark matter halos \citep{Diemer2014DependenceRate, Adhikari2014SplashbackHalos, shi16}. The splashback radius is associated with a sharp drop in the halo density profile that is created as particles pile up near the apocenters of their orbits. It has been argued that the splashback radius provides a physically motivated boundary to halos \citep{Diemer2014DependenceRate, Adhikari2014SplashbackHalos, More2015TheMass}.

Recent work has shown that a physically motivated boundary is important for understanding the properties of both galaxies and halos. For example, \citet{Baxter2017} showed that the fraction of red galaxies in {\fontfamily{lmtt}\selectfont redMaPPer} clusters \citep{Rykoff2014RedMaPPer.Catalog} displays an abrupt decrease around the location of the splashback radius. It has also been shown that assembly bias is heavily dependent on halo mass definitions \citep{More2016DetectionClusters, Villarreal2017TheBias, Chue2018SomeHalos, Mansfield2019TheBias}. Current halo finders use definitions of halo mass and halo radius based on somewhat arbitrary choices for the overdensity, $\Delta$. As such, these standard halo mass definitions do not necessarily correspond to the virialized mass of the halo. More physically motivated definitions, such as the splashback radius, can conceal discrepancies in the assembly bias measurements \citep{Chue2018SomeHalos}. It has also been suggested that the splashback radius can be used to measure dynamical friction \citep{Adhikari2016ObservingClusters} and to constrain alternative theories of gravity and self-interacting dark matter \citep{Adhikari2018SplashbackGravity, Banerjee2019SignaturesDistributions}. 

\newpage

Recent theoretical interest in the splashback radius naturally raises the question of how well it can be measured in data. Based on the spherical collapse model, it has been suggested that the splashback radius can be approximated by the minimum in the slope of the density profile of dark matter halos \citep{Adhikari2014SplashbackHalos, More2015TheMass}. However, it is a common misunderstanding that the splashback radius is simply the `dip' at the transition between the one and two-halo regime. Unlike the spherical collapse model, halos and their splashback boundaries are not spherical due to the scatter in particle apocenters \citep{Adhikari2014SplashbackHalos, Mansfield17shellfish}. Furthermore, the energy and momentum of particles at infall can affect their splashback radius. As such, the steepest slope of the density profile is not necessarily the true splashback radius \citep{Diemer2017TheAlgorithm}. Finally, systematics in optical observations of clusters can bias the location of the steepest slope \citep{busch_white_17}. However, being the closest observable, the location of the steepest slope has commonly been used as a definition of the splashback radius, especially in observations.

Previous studies have measured the splashback feature in stacked galaxy surface density profiles around massive galaxy clusters \citep{More2016DetectionClusters, umetsudiemer17, Baxter2017, Chang2017TheProfiles, shin19spl, Contigiani19spl, zurcher_more_2019} and in weak lensing measurements \citep{Chang2017TheProfiles}. The detection of the splashback radius is achieved by comparing two model fits: a model with the splashback feature as introduced in \citet[][]{Diemer2014DependenceRate} (the DK14 model) and a different `null' model without a splashback feature. 

DK14 uses the Einasto profile to describe the collapsed material (one-halo term) and a power-law profile for the infalling material (two-halo term) \citep{Gunn1972OnEvolution}. More importantly, the model includes a truncation of the Einasto profile at $r_t$, which introduces a minimum in the slope of the density profile corresponding to the splashback feature. 

The issue with the second `null' model, is that there is no physical basis for a splashback-free halo in a $\Lambda \mathrm{CDM}$ universe. The splashback feature is a natural consequence of the hierarchical formation of dark matter halos \citep{Fillmore1984Self-similarUniverse, Bertschinger1985Self-similarUniverse} so all dark matter halos should have a splashback radius. Hence, unless one is assuming a non-$\Lambda \mathrm{CDM}$ universe, there is no natural `null' model with which to compare. \textit{Instead of framing the detection issue as a model selection problem, we should be asking how precisely and accurately we can measure the splashback radius}. 

In the following paragraphs we describe the methods used by \citet{More2016DetectionClusters}, \citet{Baxter2017} and \citet{Chang2017TheProfiles} to claim detection:

\begin{enumerate}
    \item \citet{More2016DetectionClusters} were the first to claim a detection of the splashback feature in real data. They used the Sloan Digital Sky Survey (SDSS) DR8 data to measure surface density profiles around galaxy clusters using the {\fontfamily{lmtt}\selectfont redMaPPer} cluster catalog. \citet{More2016DetectionClusters} followed a model selection approach to determine the detection of the splashback feature. They defined an alternative DK14 model composed of a pure Einasto profile (setting  $f_\text{trans}= 1$) and a power-law term to describe the density profile without a splashback feature. When compared to the splashback-free model, the original DK14 provided a better fit to the data. This suggested that the data disfavored the splashback-free model, thus proving the detection of the splashback radius. \citet{More2016DetectionClusters} defined the splashback radius as the steepest slope of their best fit DK14 profile. 

    \item \citet{Baxter2017} divided the collapsed and infalling regions of the density profile and studying only the collapsed part for the detection of the splashback radius. They chose the 1-halo term of the Navarro, Frenk, and White (NFW) profile \citep{Navarro1996TheHalos} to be the null splashback-free model. They used a Bayesian approach to fit DK14, including miscentering on the same dataset as \citet{More2016DetectionClusters}. They computed the location and steepness of the steepest slope by rebuilding the density profile and its log slope from the posteriors of their free parameters. Finally, they compared the slope of the DK14 fit collapsed region to that of the NFW fit. Because the collapsed region of DK14 was steeper than that of the NFW fit, they claimed a successful detection of the splashback radius.
    
    \item \citet{Chang2017TheProfiles} detected the splashback feature around {\fontfamily{lmtt}\selectfont redMaPPer} clusters with the first year of DES data using both surface density of galaxies and weak lensing profiles. Following the same approach as \citet{Baxter2017}, \citet{Chang2017TheProfiles} demonstrated that the location and steepness of the collapsed term for the weak lensing profiles agreed with those measured from the stacked density profiles.
  
\end{enumerate}

The goal of this paper is to study the accuracy and precision to which we can measure the splashback radius. First, we study how accurately we can measure the splashback radius from 3D density and subhalo profiles. We use results from {\fontfamily{lmtt}\selectfont SPARTA} (Subhalo and PARticle Trajectory Analysis), an algorithm that tracks particle trajectories to measure the splashback radius \citep{Diemer2017TheAlgorithm, Diemer2017TheCosmologyb}. We compare the splashback radius measured with {\fontfamily{lmtt}\selectfont SPARTA} with the location of the steepest slope of the density and subhalo profiles for a given halo sample. We consider scenarios in which halos are selected not only by mass but also by secondary halo properties such as halo mass accretion rate. This choice was motivated by previous work showing that the splashback radius is most strongly correlated with accretion rate \citep{Diemer2014DependenceRate, Adhikari2014SplashbackHalos, More2015TheMass}. 

Second, we study how precisely we can measure the splashback radius from weak lensing data. Weak lensing is a direct probe of the mass profile of dark matter halos, and, thus, ideal for the detection of the splashback radius. We discuss how well we can constrain the correlation of the splashback radius with accretion rate at fixed halo mass for current surveys such as the Hyper Suprime Cam survey \citep[HSC,][]{Aihara2018FirstProgram}, and future surveys like the Large Synoptic Survey Telescope \citep[LSST,][]{Ivezic2008LargeDesign}, Euclid \citep{Laureijs2011EuclidReport} and the Wide Field Infrared Survey Telescope \citep[WFIRST, ][]{Spergel2013Wide-FieldReport}. Here, we only consider dark matter simulations without gas. We are also only studying the ideal case in which clusters have been correctly identified, without the impact of errors due to cluster finders or miscentering.

This paper is structured as follows. In Section 2, we introduce our halo and subhalo sample along with different definitions of the splashback radius. In this section, we also introduce the methods we use to compute and fit density, subhalo, and weak lensing profiles. We present and discuss our results in Sections 3 and 4. Finally, we summarize our work in Section 5. 

We adopt the cosmology of the MultiDark Planck 2 (MDPL2) simulation \citep{Prada2011HaloCosmology}, namely, a flat, $\Lambda \mathrm{CDM}$ cosmology with  $\Omega_M = 0.307$, $\Omega_b = 0.0482$, $\sigma_8 = 0.829$, $h = 0.678$, $n_s=0.9611$, corresponding to the best-fit Planck cosmology \citep{Ade2014PlanckParameters}.

\section{Methods}

\subsection{Halo and Subhalo Selections}

We aim to study cluster-sized halos similar to the sample used by \citet{Chang2017TheProfiles}. We base our analysis on the publicly available MultiDark Planck 2 (MDPL2) simulation \citep{Prada2011HaloCosmology}. We select halos from the {\fontfamily{lmtt}\selectfont ROCKSTAR} halo catalog \citep{Behroozi2013TheCores, Behroozi2013GravitationallyCosmology} at $z=0.36$. We pick host halos within a narrow mass range of $ M_{\rm 200m}=10^{13.8} \mathrm{M}_\odot/h$ to $M_{\rm 200m}=10^{14.1} \mathrm{M}_\odot/h$. Both the mass range and the redshift correspond to the best fit halo mass of {\fontfamily{lmtt}\selectfont redMaPPer} clusters with richness $20 < \lambda < 100$ and a redshift range of $[0.2, 0.55]$ in \citet{Chang2017TheProfiles}. 

Given that the location of the splashback radius is correlated with halo accretion rate \citep{Diemer2014DependenceRate, Adhikari2014SplashbackHalos, More2015TheMass}, we bin our halo sample by accretion rate. {\fontfamily{lmtt}\selectfont SPARTA} and  {\fontfamily{lmtt}\selectfont ROCKSTAR} use different definitions for accretion rate over 1 dynamical time ($\Gamma_{\rm dyn}$). Instead of using the definitions in the {\fontfamily{lmtt}\selectfont ROCKSTAR} catalogs, we recompute \mar from the {\fontfamily{lmtt}\selectfont ROCKSTAR} merger trees as in \citet{Diemer2017TheAlgorithm}: 
\noindent
\begin{equation}
    \Gamma_\mathrm{dyn} = \frac{\Delta \mathrm{log} M_{\rm 200b}}{\Delta \mathrm{log} a},
\end{equation}
\noindent
where $M_{\rm 200b}$ is the halo mass defined relative to an overdensity of $200\overline{\rho}$ where $\overline{\rho}$ is the mean matter density. Details about this process can be found in \citet{Xhakaj_2019}. {\fontfamily{lmtt}\selectfont SPARTA} and  {\fontfamily{lmtt}\selectfont ROCKSTAR} employ different definitions for halo mass. \citet{Diemer2017TheAlgorithm} measures \mass using both bound and unbound particles while {\fontfamily{lmtt}\selectfont ROCKSTAR}'s default setting measures \mass using only bound particles (see \citealt{Xhakaj_2019} for more information). Here we use the  {\fontfamily{lmtt}\selectfont ROCKSTAR} \mass definition. 

We neglect a subset (2\%) of our halo sample that has negative mass accretion rates. These could be mergers or fly-bys. After selecting host halos with $10^{13.8}<{M_{\rm 200m}}<10^{14.1} \mathrm{M}_\odot/h$ and removing halos with negative accretion rates, our final sample consists of roughly 25000 halos. 

\begin{figure*}
    \includegraphics[width=.9\textwidth]{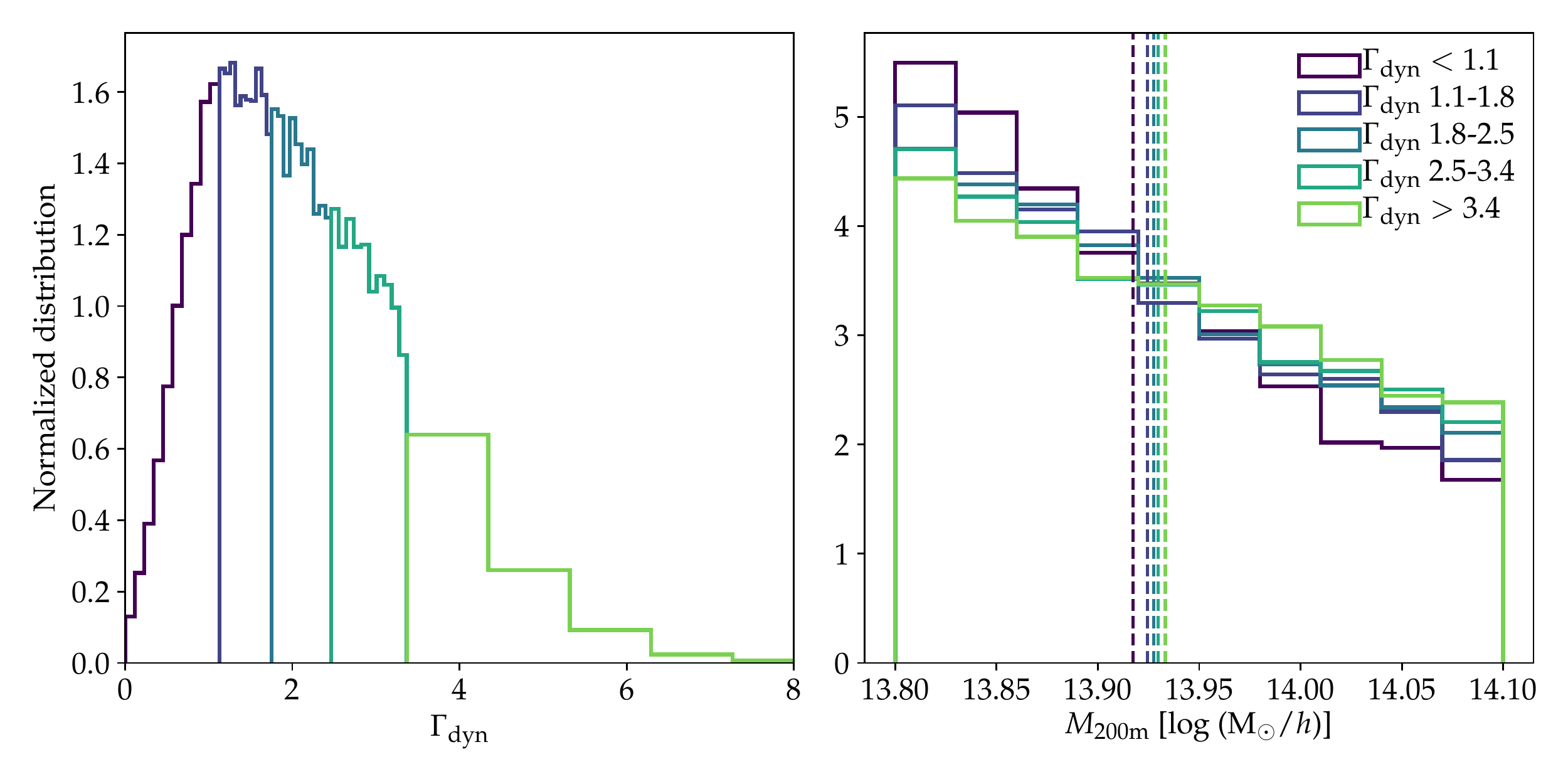}
  \caption{Left: Halo mass accretion rates of our five halo samples. Right: halo mass distributions of the same samples. Our five samples differ in their mean accretion rates but have similar mean halo masses (dashed vertical lines). Each bin has an equal number of halos and a similar mass distribution.}  \label{fig:mass_dist}
\end{figure*}

We divide this catalog by mass accretion rate such that each bin contains an equal number of about 5000 halos. Each sub-sample has a similar mass distribution (Figure \ref{fig:mass_dist}). Our accretion rate bins are $[0, 1.1, 1.8, 2.5, 3.4, 13.1]$.

We also study the 3D profiles of subhalos around our $M_{\rm 200m}=10^{14}\mathrm{M}_\odot/h$ halo sample (see Section 3.2). We consider subhalos that will host galaxies from the upcoming DESI (Dark Energy Spectroscopic Instrument) experiment \citep{DESICollaboration2016TheDesign}. The DESI Bright Galaxy Sample (BGS) is a flux-limited sample selected with an r-band magnitude threshold of 19.5 (Omar Ruiz Macias et al. {\it in prep}). We select subhalos within the mass bin $10^{12} <{M_{ \rm peak}}<10^{12.06} \mathrm{M}_\odot/h$. The number density of this sample is $\bar{n} = 10^{-3} \mathrm{Mpc}/h$ at a redshift of 0.36, which matches the expected number density of BGS. We also want the subhalo sample to reflect the Y1 area that will be covered by DESI. For this purpose, when studying subhalo profiles, we limit the volume used to extract profiles to the expected DESI Y1 area (comoving volume of $0.6~(\mathrm{Gpc}/h)^3$).

\begin{figure*}
    \includegraphics[width=\textwidth]{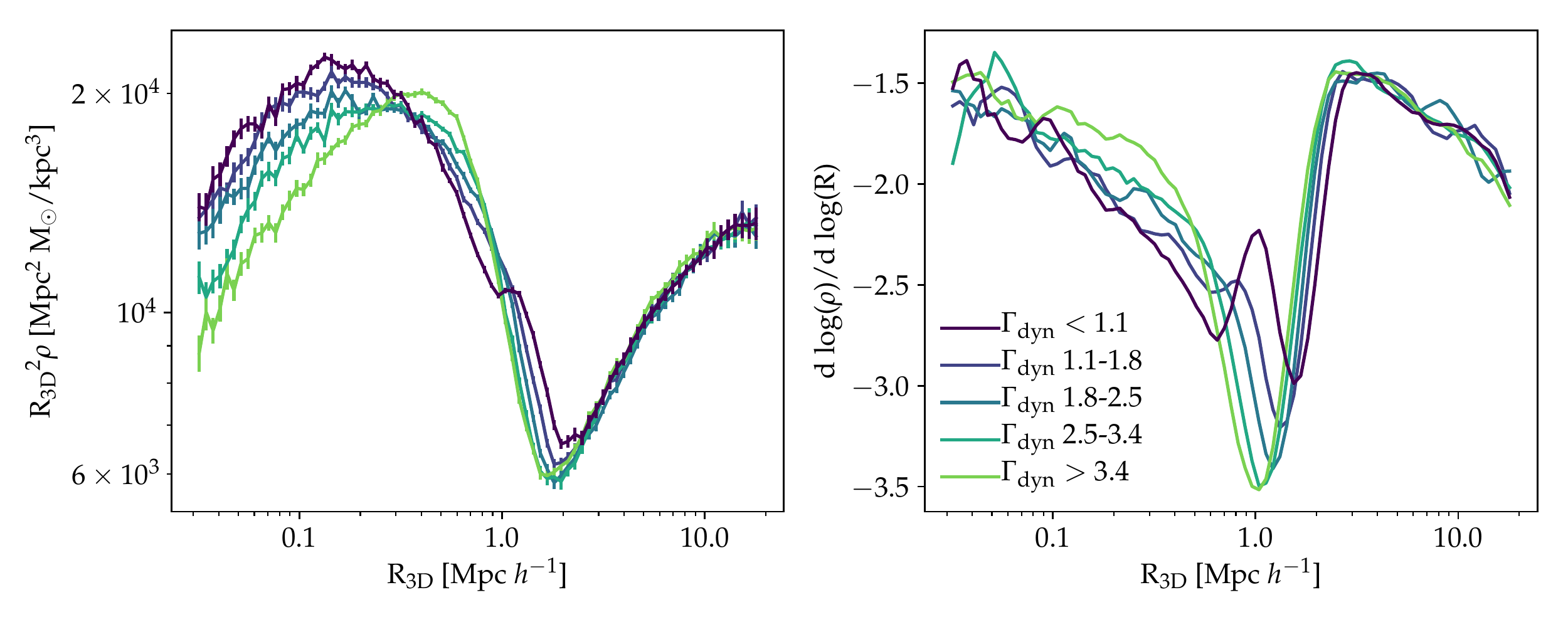}
  \caption{Left: 3D density profiles of our five samples with jackknife error bars (representing sample variance in the simulation). Each bin in accretion rate contains an equal number of halos.  Right: corresponding logarithmic slopes of the profiles as computed using the Savtzky-Golay filter. The minimum in the log slope varies with $\Gamma_{\rm dyn}$. For low accretion rate halos, the appearance of a second caustic is apparent (blue and purple lines).}
  \label{fig:all_rho_gamma}
\end{figure*}

\subsection{Splashback Radius Modeling and Definitions}

 We now introduce various definitions of the splashback radius that have been used in previous work. We also discuss how previous work has modeled and measured the splashback radius. In this paper, the splashback radius is denoted as $R_{\rm sp}$.

\subsubsection{Splashback Radius from Particle Dynamics}

{\fontfamily{lmtt}\selectfont SPARTA} measures particles' apocentric radii by tracing their trajectories and thus provides a direct measurement of the splashback radius (hereafter $R_{\rm SPARTA}$) \citep{Diemer2017TheAlgorithm, Diemer2017TheCosmologyb}. {\fontfamily{lmtt}\selectfont SPARTA} has not been run on MDPL2. To obtain {\fontfamily{lmtt}\selectfont SPARTA} based determinations of the splashback radius, we match our MDPL2 halo sample with that of the L500-Planck simulation at the same redshift (z = 0.36). L500-Planck is a 500 Mpc box simulation with a Planck-like cosmology, on which {\fontfamily{lmtt}\selectfont SPARTA} has already been run \citep{DK15, Diemer2017TheAlgorithm}. To obtain the distributions of \rsp for our sample, we make the same \mass and \mar cuts in L500-Planck as for our MDPL2 sample at z = 0.36. Particles infalling onto halos have a range of apocentric radii. In order to have a more compact definition of $R_{\rm sp}$, {\fontfamily{lmtt}\selectfont SPARTA} catalogs provide the 50th, 63rd, 75th and 87th percentiles of the splashback radius measurements from individual particles. Hereafter we abbreviate {\fontfamily{lmtt}\selectfont SPARTA}'s n$^{\rm th}$ percentile measurement of the splashback radius as $R^{\rm n}_{\rm SPARTA}$.

This matching procedure is valid because (1) we have verified that \mar is identical \citep{Xhakaj_2019}, and (2) we are interested in the statistics of \rsp in binned halo samples rather than \rsp of individual halos. The difference in resolutions between L500-Planck and MDPL2 will not affect our results, given that we are studying cluster outskirts, which are not affected by numerical artifacts.

\subsubsection{Splashback Radius from Density Profiles}

Another definition of the splashback radius is the location of the steepest slope of the density profile. This minimum in the slope roughly marks the separation of the collapsed and infalling regions of the halo. 

We compute the logarithmic slopes of the profiles, both parametrically and non-parametrically, through the DK14 model and the Savitzky-Golay (SG) method \citep{1964AnaCh..36.1627S}. The SG filter fits the profiles using a 4th order polynomial in radial bins \citep{Diemer2014DependenceRate, More2015TheMass}. The DK14 model describes the profile as comprised of 2 parts: a truncated Einasto profile, describing the collapsed region, and a power-law term, describing the infalling region of the halo: 
\begin{align}
\label{eqn:eqlabel}
\begin{split}
& \rho(r) = \rho_{\text{infall}}(r) + \rho_{\text{coll}}(r)  
\\
& \rho_{\text{coll}}(r) = \rho_{\text{Ein}}(r)f_{\text{trans}}(r)
\\
& \rho_{\text{Ein}}(r) = \rho_s \text{exp}\Big(-\frac{2}{\alpha}\Big[ \Big( \frac{r}{r_s}\Big)^\alpha - 1 \Big] \Big)
\\ 
& f_{\text{trans}}(r) = \Big[ 1 + \Big( \frac{r}{r_t} \Big)^\beta \Big]^{-\gamma/\beta}
\\
& \rho_{\text{infall}}(r) =  \rho_0 \Big( \frac{r}{r_0} \Big)^{-s_e}. 
\\
\end{split}
\end{align}
The truncation of the Einasto profile, implemented through $f_{\text{trans}}(r)$, introduces a minimum in the slope of the density profile, which accounts for the steepening at $R_{\rm sp}$. The model has 8 free parameters: $\rho_s$, the central scale density, $r_s$, the scale radius, $\alpha$, the steepening of the inner slope of the Einasto profile, $r_t$, the truncation radius, $\beta$, the sharpness of the steepening, and $\gamma$, the asymptotic negative slope of the steepening term. The location of the steepest slope measured with the SG filter and the DK14 model are hereafter abbreviated as $R_{\rm SG}$ and $R_{\rm DK14}$.

\begin{figure*}
    \includegraphics[width=.9\textwidth]{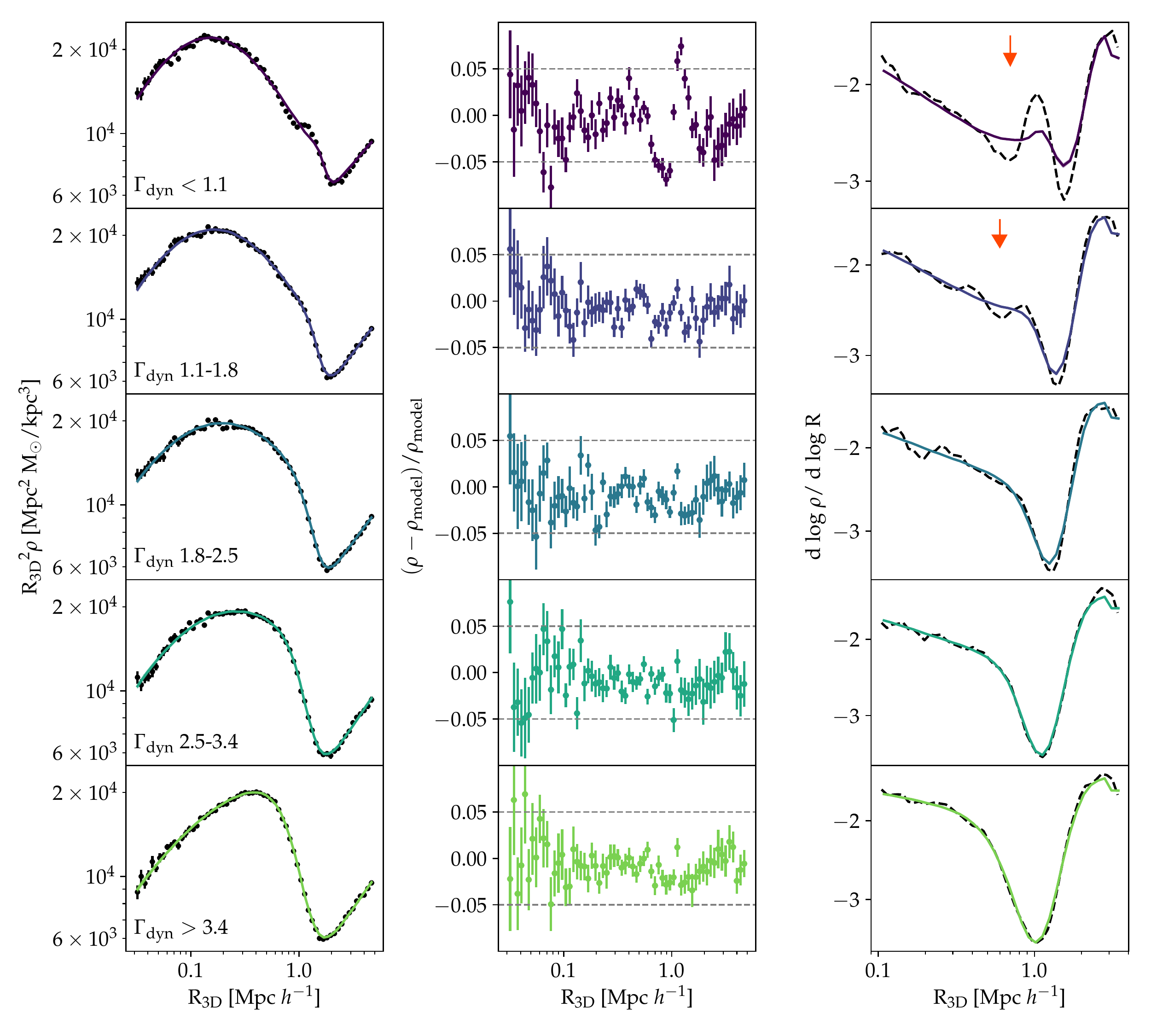}
  \caption{Left: best-fit DK14 models (colored lines) for $\rho^{\rm 3D}$ measured in MDPL2 (black dots) in bins of accretion rate. Mass accretion rate increases from top to bottom while the mass of the halo sample is constant throughout all bins (\mass$=10^{14} \mathrm{M}_\odot/h$). Middle: fractional accuracy of the DK14 model. Right: comparison between the values of the logarithmic slope as computed from the DK14 fits (colored lines) and those computed through a Savitzky-Golay filter (black lines). The accuracy of the DK14 is generally better than 5\% in agreement with \citet{Diemer2014DependenceRate}. However, the DK14 model fits do not capture the appearance of the second caustic (red arrows). We will show that this can lead to difficulties when using DK14 to fit for the minimum of the log-slope of $\rho^{\rm 3D}$ in low accretion rate bins.}
  \label{fig:model_fits}
\end{figure*}

\subsection{Fiducial Density, Suhalo, and Weak Lensing Profiles}

We use {\fontfamily{lmtt}\selectfont Halotools} \citep{Hearin2017ForwardHalotools} to compute stacked 3D density, subhalo, and weak lensing profiles for our fiducial halo sample. The density profile can be computed from the cross-correlation function as: 
\begin{equation}
\langle \rho(r) \rangle = \bar{\rho}_\text{m} (~\xi(r)_\mathrm{12}+1~),
\end{equation}

\noindent where $\bar{\rho}_m$ is the mean matter density of the universe, and $\xi(r)_\mathrm{12}$ is the two-point cross-correlation function computed with the  {\fontfamily{lmtt}\selectfont Halotools} function {\fontfamily{lmtt}\selectfont tpcf}. To compute 3D density profiles ($\rho^{\rm 3D}$) we cross-correlate host halos with dark matter particles, while for the 3D subhalo profiles ($\rho^{\rm 3D}_{\rm sub}$) we cross-correlate host halos with subhalos. 

We compute the weak lensing profile ($\Delta\Sigma$) for our fiducial halo sample by measuring the excess surface density of dark matter particles in cylinders surrounding host halos using the {\fontfamily{lmtt}\selectfont Halotools} function {\fontfamily{lmtt}\selectfont DeltaSigma}.

\subsection{Fitting Profiles with the DK14 model}

We model the 3D density, subhalo, and weak lensing profiles with DK14 using the python toolkit {\fontfamily{lmtt}\selectfont COLOSSUS}  \citep{Diemer2017COLOSSUS:Halos}. The functional form of the 3D density profile is shown in Equation 1. The projected surface mass density, $\Sigma(R)$, is the integral of the density profile along the line of sight: 
\begin{equation}
    \Sigma(R) = \int^{h_\mathrm{max}}_{-h_\mathrm{max}}  \rho \sqrt{R^2+h^2} dh,
\end{equation}

\noindent where $h_{\rm max}$ is the maximum line of sight integration length, namely $10^{8} \mathrm{kpc}/h$. The excess surface mass density,  $\Delta\Sigma(R)$, is then:
\begin{equation}
    \Delta\Sigma(R) = \frac{1}{\pi R^2} \int^R_0 2\pi r \Sigma(r) dr - \Sigma(R).
\end{equation}

Following \citet{Baxter2017} and \citet{Chang2017TheProfiles} we use a Bayesian approach to fit DK14 to $\rho^{\rm 3D}$, $\rho^{\rm 3D}_{\rm sub}$, and $\Delta\Sigma$. We adopt similar priors to \citet{Chang2017TheProfiles} (see Table \ref{table1}). 
Most of the parameters have wide uniform priors. We use Gaussian priors on $\alpha$ motivated by  \citet{gaoalpha2008}, and on $\beta$ and $\gamma$ as recommended by \citet{Diemer2014DependenceRate}. We sample the posterior parameter space with a Markov Chain Monte Carlo (MCMC) analysis implemented in {\fontfamily{lmtt}\selectfont emcee} \citep{Foreman-Mackey2012Emcee:Hammer}. We assess for convergence using trace plots and Kolmogorov-Smirnov statistic. Once the chains are converged, we rebuild the profiles using the parameters chosen by the chain in each iteration. From these model profiles, we compute the posterior distribution of $R_{\rm DK14}$ (minimum of the logarithmic slope).

\begin{table}
\centering
\begin{tabular}{@{\hskip 0.5in}lc@{\hskip 0.5in}}
\toprule
Parameter     & Priors          \\ \midrule

\addlinespace[0.15cm]
log($\rho_s$)      & $\mathcal{U}$(4, 8) log($\mathrm{M}_\odot h^2/\mathrm{kpc}^{3}$) \\
\addlinespace[0.15cm]
log($r_t$)         & $\mathcal{U}$(2, 4) log($\mathrm{kpc}/h$)                \\
\addlinespace[0.15cm]
log($r_s$)         & $\mathcal{U}$(0, 4) log($\mathrm{kpc}/h$) \\
\addlinespace[0.15cm]
log($\alpha$) & $\mathcal{N}$(-0.67, 0.16)            \\
\addlinespace[0.15cm]
$\beta$  & $\mathcal{N}$(4, 0.1)                    \\
\addlinespace[0.15cm]
$\gamma$ & $\mathcal{N}$(6, 0.1)   \\
\addlinespace[0.15cm]
log($\rho_0$)      & $\mathcal{U}$(0, 0.5)  log($\mathrm{M}_\odot h^2/\mathrm{kpc}^{3}$) \\
\addlinespace[0.15cm]
$s_e$         & $\mathcal{U}$(1, 10)          \\    
\bottomrule
\end{tabular}
\caption{Priors used to fit DK14 to the 3D density and weak lensing profiles.}
\label{table1}
\end{table}

\subsection{Summary of Notation}

To summarize, throughout this paper, we use the following notation for different \rsp measurements. 

\begin{itemize}
    \item $R^{50}_\mathrm{SPARTA}$: 50th percentile  of \rsp measured with {\fontfamily{lmtt}\selectfont SPARTA} from the particle trajectories of L500-Planck and matched with our halo sample (see Section 2.2.1). Similarly, $R^{67}_\mathrm{SPARTA}$ denotes the 67th percentile and so on and so forth.
    \item $R_\mathrm{SG}$: location of the steepest slope of $\rho^{\rm 3D}$ measured with SG. 
    \item $R_\mathrm{DK14}$: location of the steepest slope of $\rho^{\rm 3D}$ measured with DK14.
    \item $R_\mathrm{sub}$: location of the steepest slope of $\rho^{\rm 3D}_{\rm sub}$. 
    \item $R_\mathrm{SG, sub}$: location of the steepest slope of $\rho^{\rm 3D}_{\rm sub}$ measured with SG. 
    \item $R_\mathrm{DK14, sub}$: location of the steepest slope of $\rho^{\rm 3D}_{\rm sub}$ measured with DK14.
\end{itemize}

\section{Results}

We now examine how \rsp estimates computed from density, weak lensing, and subhalo profiles compare to the distribution of \rsp values measured with {\fontfamily{lmtt}\selectfont SPARTA}. 

\subsection{Splashback Radius Estimate from 3D Density Profiles}

We measure the 3D density profiles for our fiducial sample in bins of accretion rate (Figure \ref{fig:mass_dist}) following the methodology described in Section 2.3. The profiles are shown in Figure \ref{fig:all_rho_gamma}, which displays $\rho^{\rm 3D}$ and the log-slopes of $\rho^{\rm 3D}$ for halos binned by accretion rate. Figure \ref{fig:all_rho_gamma} shows that the minimum of the log-slopes of $\rho^{\rm 3D}$ shifts to smaller scales and becomes deeper with increasing $\Gamma_{\rm dyn}$. For \mar$<1.8$, however, a second minimum is apparent at radii smaller than the splashback radius. This feature is a second caustic: a second sharp drop in $\rho^{\rm 3D}$. Caustics are caused by particle orbits that pile up at the same location at the apocenters of their orbits \citep{Adhikari2014SplashbackHalos}. The second caustic corresponds to the location where particles reach the apocenter of their second orbit. The trends in Figure \ref{fig:all_rho_gamma} are in agreement with results from previous work \citep{Diemer2014DependenceRate, Adhikari2014SplashbackHalos, More2015TheMass, Diemer2017TheCosmologyb}. 

\begin{figure*}
    \includegraphics[width=.82\textwidth]{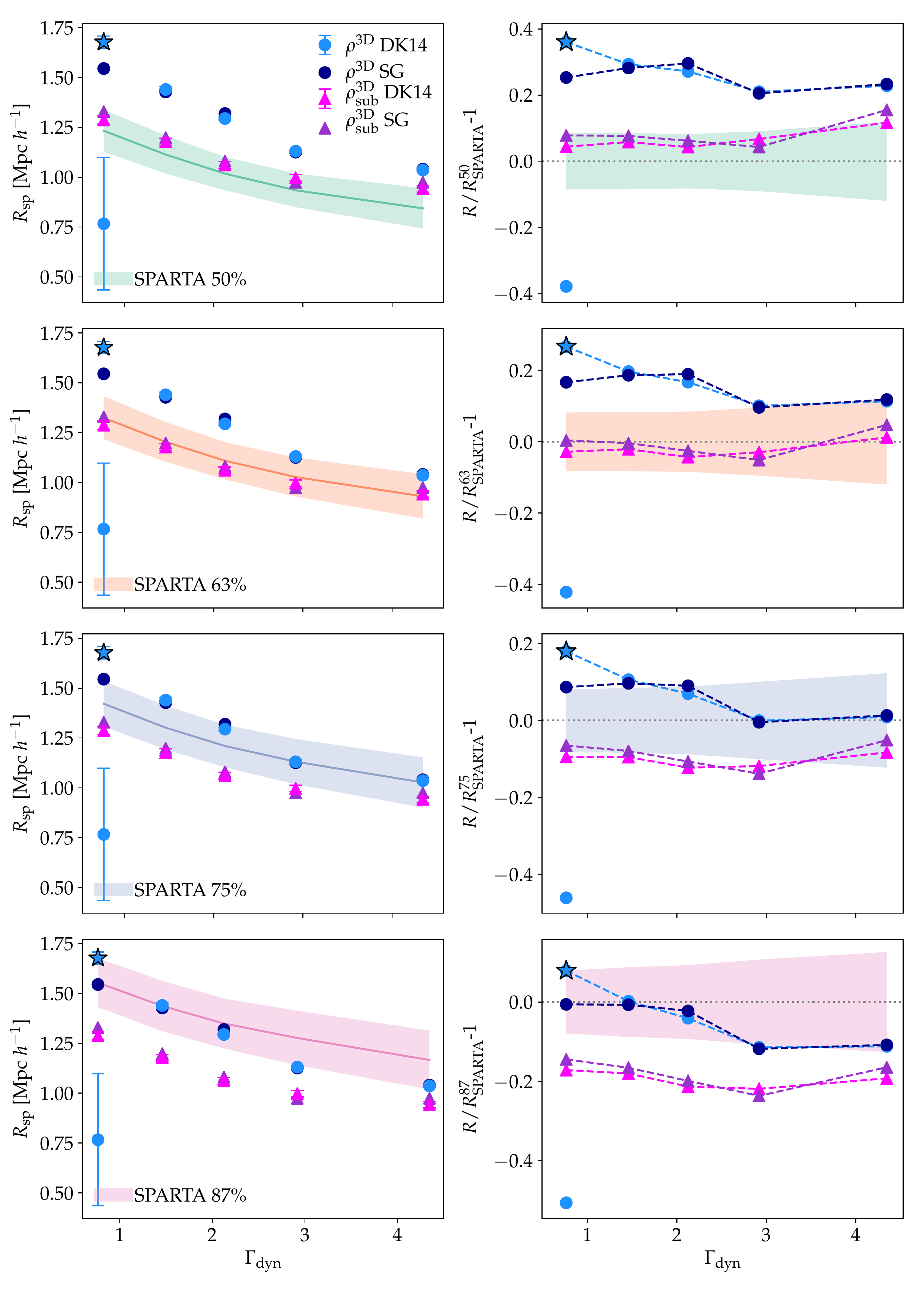}
  \caption{Left: \rsp estimates from $\rho^{\rm 3D}$ (circles) and  $\rho^{\rm 3D}_{\rm sub}$ (triangles) using both the DK14 model fits and Savitzky-Golay. Rows from top to bottom compare these \rsp estimates with different percentiles of $R_\mathrm{SPARTA}$ (shaded regions). Solid lines represent the mean value of the percentile for each bin in \mass and $\Gamma_{\rm dyn}$. The width of the shaded region is the $1\sigma$ spread of percentile values. The blue star displays the value of \rsp for the lowest accretion rate bin when selecting the higher peak in the bimodal distribution of the posterior (Figure \ref{fig:rsp_poster}). Right: ratio of estimated \rsp values to different percentiles of $R_{\rm SPARTA}$. The steepest slope of $\rho^{\rm 3D}$ does not correspond to any single percentile values from {\fontfamily{lmtt}\selectfont SPARTA}. Instead, it matches $R^{75}_{\rm SPARTA}$ for the higher accretion bins (\mar > 2.5), and  $R^{87}_{\rm SPARTA}$ for the lower accretion bins (\mar < 2.5). On the other hand, the steepest slope of $\rho^{\rm 3D}_{\rm sub}$ matches $R^{63}_{\rm SPARTA}$ consistently across all \mar bins. Finally, DK14 fails in the low accretion regime. This is due to the appearance of the second caustic in the density profile, which causes a bimodal posterior distribution.}
  \label{fig:rsp_steepest_slope}
\end{figure*}

We fit the 3D density profiles in Figure \ref{fig:all_rho_gamma} with DK14 and SG and display the results in Figure \ref{fig:model_fits}. \citet{Diemer2014DependenceRate} showed that DK14 fits the data with a fractional accuracy of 5 to 10\%. Figure \ref{fig:model_fits} shows that indeed, the accuracy of the DK14 fits is, in general, better than 5\%. However, the model does not capture the appearance of the second caustic that arises in the lower accretion rate bins (\mar < 1.8). 

Finally, in Figure \ref{fig:rsp_steepest_slope}, we study how \rsp estimated from 3D density profiles compares to the apocentric radius of particles computed with {\fontfamily{lmtt}\selectfont SPARTA}. Particles infalling onto halos will have a range of apocentric radii. This means that a given halo will not have a single fixed \rsp, but rather a \emph{distribution} of values. For this reason, {\fontfamily{lmtt}\selectfont SPARTA} provides the 50th, 63rd, 75th and 87th percentile of the apocenteric radius of individual particles. Each of our bins in \mass and \mar is comprised of a sample of halos. Therefore, each bin in \mass and \mar will have a distribution of values for a given percentile. This distribution is indicated by the shaded regions in Figure \ref{fig:rsp_steepest_slope}. For example, the upper left panel of Figure \ref{fig:rsp_steepest_slope} compares $R_{\rm DK14}$ and $R_{\rm SG}$ with $R^{50}_{\rm SPARTA}$. The width of the green shaded region represents the $1\sigma$ distribution of $R^{50}_{\rm SPARTA}$, and the solid green line in the middle is the mean value of $R^{50}_{\rm SPARTA}$ in each accretion rate bin.

Figure \ref{fig:rsp_steepest_slope} shows that the steepest slope of $\rho^{\rm 3D}$ agrees with $R^{75}_\mathrm{SPARTA}$ for the higher accretion bins (\mar>2.5), and with $R^{87}_\mathrm{SPARTA}$ for the lower accretion bins (\mar<2.5).  Furthermore, $R_{\rm SG}$ and $R_{\rm DK14}$ agree throughout all accretion bins except in the lowest one (\mar<1.1). Unlike SG, DK14 largely underestimates the steepest slope for the lowest accretion bin. This is due to the second caustic that becomes apparent in the lower accretion bins (\mar<1.8) and is most prominent in the lowest bin. Thus, using the steepest slope as the \rsp estimate for halos with low accretion rates will lead to biased measurements. 

\subsection{Splashback Radius Estimate from 3D Subhalo Profiles}

Given that in observations, we measure satellite (and thus subhalo) profiles, we now investigate how the steepest slope of $\rho^{\rm 3D}_{\rm sub}$ varies with accretion rate. We focus in particular on subhalos that roughly correspond to a DESI-like selection (see Section 2.1). We measure $R_{\rm sub}$ by fitting the subhalo profiles with DK14 and SG. Errors are computed by resampling over host halos. Figure \ref{fig:rsp_steepest_slope} compares $R_{\rm sub}$ estimated from the steepest slope of $\rho^{\rm 3D}_{\rm sub}$ to \rsp estimated from particle apocenters. Figure \ref{fig:rsp_steepest_slope} conveys that $R_{\rm sub}$ agrees with $R^{63}_\mathrm{SPARTA}$ across all accretion bins. This shows that $R_{\rm sub}$ is not susceptible to fitting artifacts due to the second caustic. Additionally, $R_{\rm sub}$ does not trace the steepest slope of the 3D density profiles. The systematically lower $R_{\rm sub}$ may indicate evidence of the dynamical friction drag due to the massive subhalos in our sample \citep{Adhikari2016ObservingClusters}. 

\subsection{Splashback Radius Estimate from Weak Lensing}

\begin{figure}
    \includegraphics[width=\columnwidth]{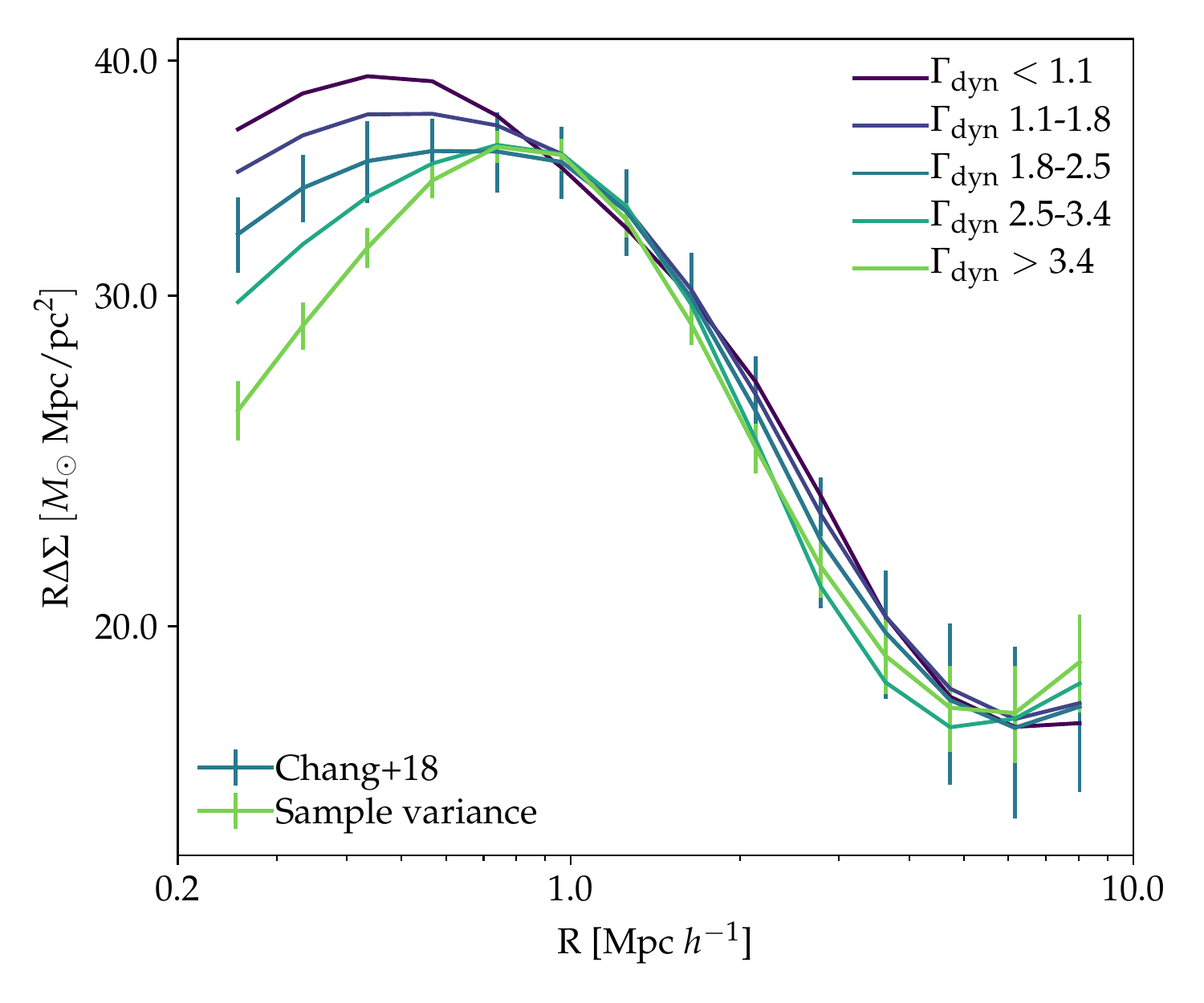}
  \caption{Weak lensing profiles for the same halo samples as in Figure \ref{fig:all_rho_gamma}. Blue error bars correspond to the errors reported by \citet{Chang2017TheProfiles} for the DES Y1 data. Green error bars are computed by resampling over host halos in MDPL2.}
  \label{fig:ds_all_gamma}
\end{figure}

Gravitational lensing is potentially the most direct method for detecting \rsp since it traces the mass profile of dark matter halos and will not be affected by issues such as dynamical friction that can bias $R_{\rm sub}$. However, weak lensing only measures the projection of $\rho^{\rm 3D}$. We expect projection effects to wash out the splashback feature, making the minimum of the logarithmic slope around \rsp broader and, therefore, harder to constrain. Figure \ref{fig:ds_all_gamma}, displays the weak lensing profiles of our fiducial halo mass sample binned by accretion rate using the same bins as in Figure \ref{fig:all_rho_gamma}. 

\begin{figure*}
    \includegraphics[width=.9\textwidth]{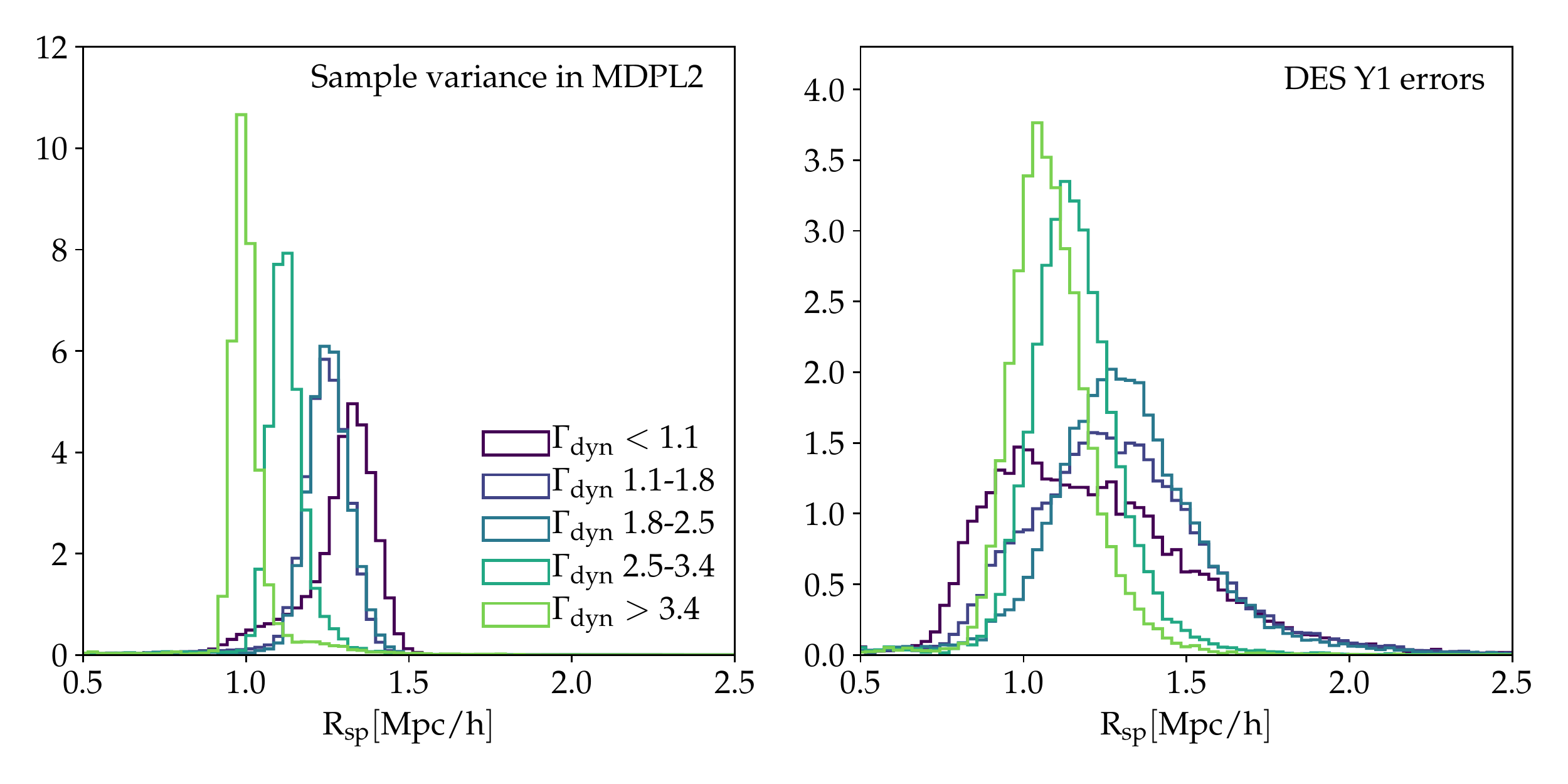}
  \caption{Steepest slope posteriors for $\Delta\Sigma$ measured with DK14. Posteriors on the left panel are computed using error bars corresponding to sample variance of the MDPL2 simulation, while those on the right are constrained with DES Y1 error bars. Different colors represent different accretion rate bins as in Figure \ref{fig:all_rho_gamma}. Although $\rho^{\rm 3D}$ and $\Delta\Sigma$ are built from the same halo sample, the location of the steepest slope for $\Delta\Sigma$ is less constrained than for $\rho^{\rm 3D}$ (Figure \ref{fig:rsp_poster}). This is because of the projection effects introduced when measuring $\Delta\Sigma$. The right-hand panel shows that the DES Y1 lensing errors are too large to allow for the detection of the \rsp - \mar connection. Conversely, this also means that DES Y1 constraints on \rsp will not be affected by any possible \mar selection effects.}
  \label{fig:post_ds_gamma}
\end{figure*}

We fit the fiducial weak lensing profiles from MDPL2 with DK14 using, in one case, the sample variance errors of the MDPL2 simulation, and in the other case, the observational error bars from reported by \citet{Chang2017TheProfiles}. Figure \ref{fig:post_ds_gamma} shows \rsp posteriors for weak lensing profiles using MDPL2 sample variance errors (left panel) and DES Y1 error bars (right panel). Profiles are color-coded by accretion rate as in Figure \ref{fig:ds_all_gamma}. Posteriors in both panels overlap with each other. Although both the 3D density and weak lensing profiles are built from the same samples (Figure \ref{fig:mass_dist}), the \rsp correlation with accretion rate is less constrained in the weak lensing profiles. This is due to the projection effects introduced when computing $\Delta\Sigma$ in projected space. The right panel shows that \rsp posteriors for the lensing profiles modeled with DES Y1 error bars are even less constrained than those with jackknife resampling. Because observational error bars are naturally larger than the jackknife ones, they put worse constraints on the posteriors.

\section{Discussion}

\subsection{Accuracy of \rsp with Estimators}

\begin{figure}
    \includegraphics[width=\columnwidth]{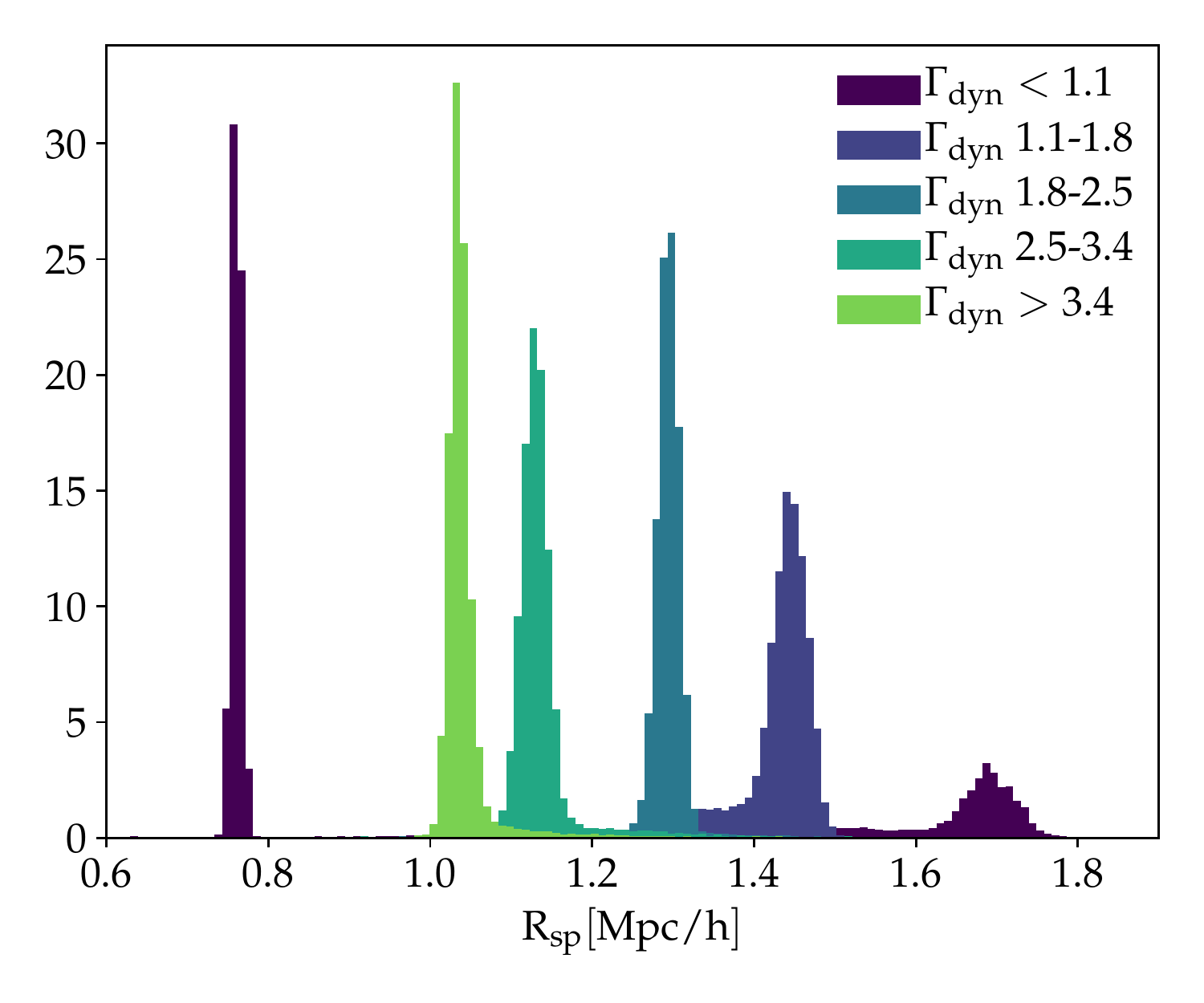}
  \caption{Posteriors of the steepest slope of $\rho^{\rm 3D}$ measured with DK14. Different colors represent different bins in accretion rate, corresponding to the same color scheme as in Figure \ref{fig:all_rho_gamma}. The lowest accretion bin (\mar <1.1) displays a bimodal distribution of the steepest slope posterior.}
  \label{fig:rsp_poster}
\end{figure}

In Figure \ref{fig:rsp_steepest_slope} we compared \rsp from particle dynamics to the location of the steepest slope in 3D density and subhalo profiles. For the density profiles, the location of the steepest slope for the higher accretion bins (\mar > 2.5) converges with $R^{75}_{\rm SPARTA}$, while for the lower accretion bins (\mar < 2.5), it converges with $R^{87}_{\rm SPARTA}$. Moreover, we find that the SG method and the DK14 profile fitting routine provide consistent results for all accretion bins, except for \mar < 1.1. Posteriors of \rsp estimates in 3D density profiles in Figure \ref{fig:rsp_poster} show that the lowest accretion bin has a bimodal distribution. All of these effects are due to the second caustic apparent in the logarithmic density slopes of slowly accreting halos, which is most clearly seen in the lowest accretion bin (\mar < 1.1). As this feature becomes evident, the location of the steepest slope matches a higher percentile of $R_{\rm SPARTA}$. Because the second caustic is not modeled by DK14, the \rsp posterior for the lowest accretion bin in Figure \ref{fig:rsp_poster} is bimodal. The first peak is located around 0.7 Mpc$/h$, while the second around 1.7 Mpc$/h$. The two modes of the posterior correspond to the first and second caustic we see in the logarithmic slope of the lowest accretion bin (Figure \ref{fig:model_fits}). However, only the one located at 1.7 Mpc$/h$ corresponds to the steepest slope of the density profile. In conclusion, the location of the steepest slope of the density profile does not correspond to a single percentile of $R_{\rm SPARTA}$. Instead, the matched $R_{\rm SPARTA}$ percentile changes with accretion rate due to the appearance of the second caustic. 

Figure \ref{fig:rsp_steepest_slope} also shows that we will be able to detect \rsp through 3D subhalo profiles with DESI Y1. When subhalos are used to trace the halo profile, we find that the \rsp estimated from subhalo profiles is smaller than that from 3D density profiles. The location of the steepest slope in subhalo profiles in Figure \ref{fig:rsp_steepest_slope} matches $R^{63}_{\rm SPARTA}$ consistently in all accretion bins. This may be due to the dynamical friction drag of the most massive subhalos in our sample. It is well known that dynamical friction causes the most massive subhalos to sink to the center of the host. \citet{More2016DetectionClusters} and \citet{Adhikari2016ObservingClusters} showed that this effect also translates into the apocentric radii of subhalos. They concluded that \rsp from subhalos for a massive subhalo sample is smaller than \rsp from particles for the same host halo. Figure \ref{fig:rsp_steepest_slope} shows a similar qualitative effect as \citet{More2016DetectionClusters} and \citet{Adhikari2016ObservingClusters}. Given that the mass of the subhalos in our sample is greater than 1\% of the host mass, the dynamical friction drag is significant in our $R_{\rm sub}$ measurements \citep{Adhikari2016ObservingClusters}. However, a more rigorous study would be required to prove that dynamical friction is indeed the primary cause of the offset seen in Figure \ref{fig:rsp_steepest_slope}. 

\begin{figure}
    \includegraphics[width=\columnwidth]{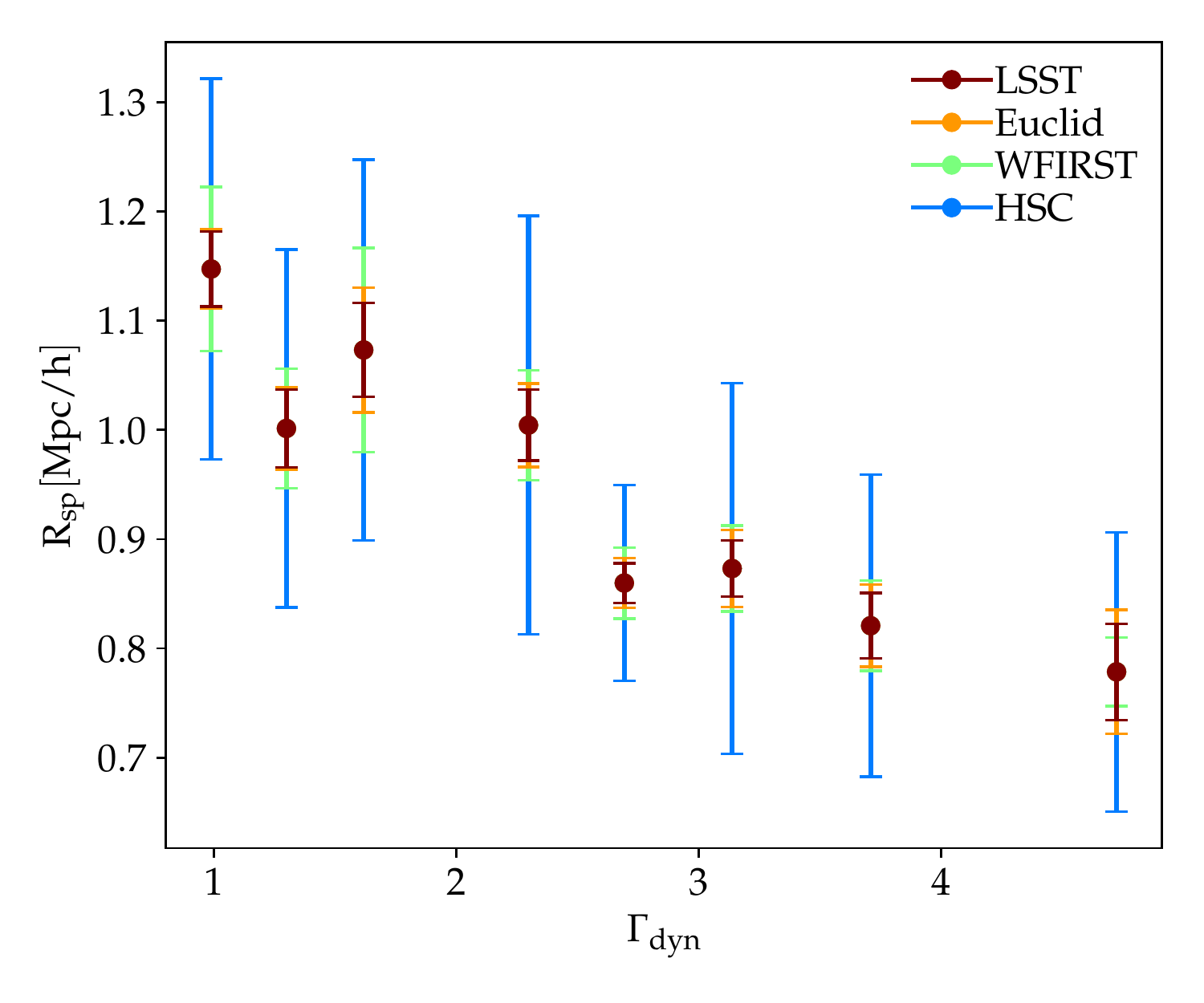}
  \caption{Splashback-accretion rate relation when considering forecast weak lensing error bars from future surveys. LSST and Euclid give the best constraint for the correlation between \rsp and $\Gamma_{\rm dyn}$.}
  \label{fig:future_surveys}
\end{figure}

\subsection{On Detecting the Correlation Between \rsp and \mar}

One of the most intriguing potential applications of the splashback radius is to observe the accretion rate of halos. The correlation between \rsp and \mar has already been proven theoretically \citep{Adhikari2014SplashbackHalos, shi16} and studied in simulations \citep{Diemer2014DependenceRate, More2015TheMass, Diemer2017TheAlgorithm, Diemer2017TheCosmologyb}. More importantly, this correlation is based on basic gravitational physics and should, in principle, be detectable in real data, if tracers of not only halo mass but also \mar can be established. However, current DES Y1 weak lensing errors are too large to constrain the \rsp - \mar trend (see Figure \ref{fig:post_ds_gamma}). The reason why this is challenging is that lensing measures a projected quantity which washes out the splashback feature. Furthermore, the signal itself is intrinsically noisy and introduces more uncertainty in the \rsp measurement than in the case of density profiles. This raises the need for better data to constrain the \mar-\rsp relation with weak lensing profiles. 

Here we assume a perfect (zero scatter) observational tracer of halo mass and accretion rate. Future work will discuss noisy tracers and their impact on the detectability of the \rsp - \mar trend. Figure \ref{fig:future_surveys} shows how precisely and accurately one can measure the splashback radius in bins of \mar for HSC, and future surveys such as LSST, Euclid, and WFIRST. We compute the lensing error bars for each of these surveys using the same methodology as the one introduced in \citet{Singh2017Galaxy-galaxyProperties} and applied in \citet{Leauthaud2019Deep+WideGalaxies}. Forecast error bars include all the terms needed to describe a Gaussian covariance and assume a perfect observational tracer of accretion rate. We do not account for effects due to selection or survey masks. Furthermore, the forecast error bars do not account for the non-gaussian covariance, and hence the signal to noise they predict is somewhat overestimated. The halo bin considered is the same as throughout the rest of this paper, namely  $10^{13.8}-10^{14.1} \mathrm{M_\odot}/h$. 

In order to study how well we can constrain the \rsp-\mar correlation in future surveys, we perform a linear fit of the data points in Figure \ref{fig:future_surveys}. The best fit slopes are $-0.10 \pm 0.02$ for HSC, $-0.11 \pm 0.02$ for WFIRST, $-0.09 \pm 0.01$ for Euclid, and $-0.09 \pm 0.01$ for LSST. All the slopes have a significance higher than $5\sigma$, with Euclid and LSST giving the tightest constraint. Thus, upcoming weak lensing surveys may be able to constrain the \mar-\rsp relation. However, this assumes a perfect observational tracer of $\Gamma_{\rm dyn}$. Further work will be necessary to construct and characterize observational tracers of $\Gamma_{\rm dyn}$.

\section{Summary and Conclusions}

In this paper, we have studied the accuracy and precision to which we can detect the splashback radius in simulated 3D density, subhalo, and weak lensing profiles. Our main goals are to (1) study how well the location of the steepest slope compares with the splashback radius from particle dynamics and (2) how precisely we can detect the splashback radius with weak lensing data given current and future surveys. We use the MDPL2 simulation to build fiducial density, subhalo, and weak lensing profiles binned by halo mass accretion rate. We measure the steepest slope parametrically, through the DK14 model, and non-parametrically, through the SG algorithm. Finally, we compare these measurements with \rsp from particle dynamics measured with  {\fontfamily{lmtt}\selectfont SPARTA}. Our main conclusions are the following: 

\begin{enumerate}
    \item The steepest slope from 3D density profiles does not agree with a single percentile of particle apocenters as measured by {\fontfamily{lmtt}\selectfont SPARTA}. The steepest slope roughly corresponds to $R^{87}_{\rm SPARTA}$ at low accretion rates and $R^{75}_{\rm SPARTA}$ at high accretion rates.
    \item For halo samples with \mar < 1.1, DK14 predicts a bimodal distribution of the steepest slope when considering the 3D density profile. This is because of the second caustic that appears in the density profiles of slowly accreting halos.
    \item It will be possible to detect \rsp using a DESI Y1-like subhalo selection through 3D subhalo profiles. 
    \item \rsp estimates from 3D subhalo profiles match $R^{63}_{\rm SPARTA}$ and are smaller than the \rsp estimates from 3D density profiles across all accretion bins. This might be due to the dynamical friction drag of the massive subhalos in our sample. 
    \item We cannot constrain the \rsp-\mar trend with DES Y1 errors. However, given an ideal observable tracer of accretion rate (zero scatter), we will be able to detect the \rsp-\mar trend with other current and future surveys such as HSC, WFIRST, Euclid and LSST. Euclid and LSST will provide the best constraints on this relation. 
\end{enumerate}{}

There is an exciting possibility that upcoming weak lensing surveys may be able to constrain the \rsp-\mar relation. Further work will be required to construct and characterize observational tracers of $\Gamma_{\rm dyn}$.

\section*{Acknowledgements}

This research was supported in part by the National Science Foundation under Grant No. NSF PHY-1748958. This material is based on work supported by the UD Department of Energy, Office of Science, Office of High Energy Physics under Award Number DE-SC0019301. AL acknowledges support from the David and Lucille Packard Foundation, and the Alfred .P Sloan foundation. EX acknowledges the generous support of Mr. and Mrs. Levy via the LEVY fellowship.

\bibliographystyle{mnras}
\bibliography{mnras_template}


\end{document}